\documentstyle[aps,multicol,epsfig]{revtex}
\newcommand{\bm}[1]{\mbox{\boldmath $#1$}}
\begin{document}

\title{Lagrangian Statistics and Temporal Intermittency in a
Shell Model of Turbulence}

\author{G. Boffetta, F. De Lillo and S. Musacchio}
\address{Dipartimento di Fisica Generale and INFM, Universit\`a di Torino \\
         Via Pietro Giuria 1, 10125 Torino, Italy}
\address{CNR-ISAC - Sezione di Torino \\
         Corso Fiume 4, 10133 Torino, Italy}
\date{\today}

\maketitle

\begin{abstract}
We study the statistics of single particle Lagrangian velocity in 
a shell model of turbulence.
We show that the small scale velocity fluctuations are
intermittent, with scaling exponents connected to the
Eulerian structure function scaling exponents.
The observed reduced scaling range is interpreted as a
manifestation of the intermediate dissipative range, as it 
disappears in a Gaussian model of turbulence.
\end{abstract}

\begin{multicols}{2}  

In the recent years there has been a great improvement in the 
experimental investigation of turbulence from a Lagrangian
point of view \cite{VSB98,OM00,LVCAB01,MMMP01}.
In the Lagrangian approach the 
flow is described by the (Lagrangian) velocity ${\bm v}({\bm x}_0,t)$
of a fluid particle initially at position ${\bm x}(0)={\bm x}_0$.
This is the natural description for studying transport and
mixing of neutrally advected substances in turbulent flows.
 
One of the simplest statistical quantities one can be interested in is 
single particle velocity increments 
$\delta {\bm v}(t)={\bm v}(t)-{\bm v}(0)$ (where, assuming
statistical homogeneity,
we have dropped the dependence on ${\bm x}_0$) for which
dimensional analysis in fully developed turbulence predicts
\cite{MY75}
\begin{equation}
\langle \delta v_{i}(t) \delta v_{j}(t) \rangle =
C_{0} \varepsilon t \delta_{ij}
\label{eq:1}
\end{equation}
where $\varepsilon$ is the mean energy dissipation and $C_{0}$ is
a numerical constant. The remarkable coincidence that the
variance of $\delta {\bm v}(t)$ grows linearly with time
is the physical basis on which stochastic models of particle
dispersion are based. It is important to recall that the
``diffusive'' nature of (\ref{eq:1}) is purely incidental:
it is a direct consequence of Kolmogorov scaling in the inertial
range of turbulence and is not directly related to a diffusive
process.
Let us recall briefly the argument leading
to the scaling in (\ref{eq:1}). We can think at the velocity
$v(t)$ advecting the Lagrangian trajectory as the superposition
of the different velocity contributions coming from the turbulent eddies
(which also move with the same velocity of the Lagrangian trajectory). 
After a time $t$ the components associated to the smaller
(and faster) eddies, below a certain scale $\ell$ are decorrelated
and thus at the leading order one has
$\delta v(t) \simeq \delta v(\ell)$. Within Kolmogorov scaling,
the velocity fluctuation at scale $\ell$ is given by
$\delta v(\ell) \sim V (\ell/L)^{1/3}$ where $V$ 
represents the typical velocity at the largest scale $L$.
The correlation time of $\delta v(\ell)$ scales as
$\tau(\ell) \sim \tau_0 (\ell/L)^{2/3}$ and thus one obtains 
the scaling in (\ref{eq:1}).

Equation (\ref{eq:1}) can be generalized to higher order moments
with the introduction of a set of temporal scaling exponent $\xi(p)$
\begin{equation}
\langle \delta v(t)^{p} \rangle \sim V^p (t/\tau_0)^{\xi(p)}
\label{eq:2}
\end{equation}
The dimensional estimation sketched above gives the prediction $\xi(p)=p/2$
but one expects corrections to dimensional scaling in presence of
intermittency.

A generalization of (\ref{eq:1}) which takes into
account intermittency corrections can be easily developed within
the multifractal model of turbulence \cite{PF85,Frisch95}.
The dimensional argument is repeated for the local
scaling exponent $h$, giving $\delta v(t) \sim V (t/\tau_0)^{h/(1-h)}$.
Integrating over the $h$ distribution one ends with the prediction
\begin{equation}
\langle \delta v(t)^{p} \rangle \sim V^p 
\int dh \left({t \over \tau_0} \right)^{[ph-D(h)+3]/(1-h)}
\label{eq:3}
\end{equation}
where, in the limit $t/\tau_0 \to 0$, the integral can be estimated
by a steepest descent argument giving finally
\begin{equation}
\xi(p)=\min_{h}\left[{ph-D(h)+3 \over 1-h} \right]
\label{eq:4}
\end{equation}
The fractal dimension $D(h)$ is related to the Eulerian structure function
scaling exponents $\zeta(q)$ by the Legendre transform
\cite{Frisch95} $\zeta(q)=\min_{h}[qh-D(h)+3]$.
The standard inequality in the multifractal model (following from 
the exact result $\zeta(3)=1$), $D(h)\le 3h+2$ implies for (\ref{eq:4})
that even in presence of intermittency $\xi(2)=1$. This is a direct
consequence of the fact that energy dissipation enters in (\ref{eq:1})
at the first power.
Our expression for scaling exponents (\ref{eq:4}) recovers in a more
compact form the prediction obtained on the basis of an ``ergodic
hypothesis'' of the statistics of energy dissipation \cite{Borgas93}.

Recent experimental results \cite{MMMP01} have shown that indeed 
Lagrangian velocity fluctuations are intermittent and display 
anomalous scaling exponents, as predicted by the above argument. 
Despite the relative high Reynolds number of the experiments, a
true temporal scaling range is not observed. Thus the estimation
of the scaling exponents in (\ref{eq:2}) can be done only
relatively to a reference moment (the so-called ESS procedure
\cite{BCTBMS93}).

In this work we use a dynamical toy model of turbulence for
investigating scaling (\ref{eq:2}) and prediction (\ref{eq:4})
at very high Reynolds numbers. The model is a popular shell model
of turbulence \cite{BJPV98} in which the velocity fluctuations
of the eddies at the scale $\ell_{n}=L 2^{-n}$ are represented by a
single variable $u_{n}$ ($n=1,2,...,N$). 
Only local interaction among shells
are represented and therefore no sweeping effects are present.
In this sense shell models are dynamical models of velocity
fluctuations in a Lagrangian framework, and have been already
used for the study of turbulent dispersion \cite{BCCV99}.
The equation of motion for the complex shell variable $u_n$ is \cite{sabra}
\begin{eqnarray}
\frac{d u_n}{dt}&=&i k_n \left( u_{n+2}u_{n+1}^*
 - {\delta \over 2} u_{n+1}u_{n-1}^*+ 
{1-\delta \over 4} u_{n-1}u_{n-2} \right) \nonumber \\
&&-\nu k_n^2  u_n +f_n
\label{eq:5}
\end{eqnarray}
where $k_n=\ell_n^{-1}$ and $f_n$ is a deterministic forcing 
acting on the first two shells only.
Shell model (\ref{eq:5}) is characterized by a chaotic dynamics
with a statistically steady state with a constant flux of energy from large 
to small scales.
The fluctuations generate by chaoticity induce a breaking of the
global scaling invariance and corrections to the Kolmogorov exponents
for the structure functions close to the experimental values \cite{BJPV98}.

Lagrangian velocity in the shell model framework can be represented
as the superposition of the contributions of all the different eddies. 
Let us thus define 
\begin{equation}
v(t) \equiv \sum_{n=1}^{N} Re(u_n)
\label{eq:6}
\end{equation}
where we have taken, rather arbitrarily, only the real part of the 
shell variables with unit coefficient. From the definition of the
shell model, there is not a precise recipe for reconstructing the
Lagrangian velocity. 
More in general, one could think of a representation in which 
shell variables are multiplied by an appropriate wavelet functions.
Of course numerical prefactor, such as $C_0$ in (\ref{eq:1}) will 
depend on the wavelet basis. Nevertheless one expects that different
choices should not affect Lagrangian scaling exponents $\xi(p)$
which are determined by the dynamical properties of the model.

Previous studies of multi-time correlations in shell models of
turbulence have demonstrated the existence of a set of correlation
times compatible with the multifractal picture of the turbulent 
cascade \cite{BBCT99}. This is an indication that, as we will see, 
Lagrangian velocity defined as (\ref{eq:6}) will be affected by intermittency.

Very long and accurate numerical simulations of the shell model (\ref{eq:5})
with $N=24$ shells and $\delta=1/2$ have been performed. 
The energy is injected at a constant flux $\varepsilon=0.01$ is the
first $2$ shells and is removed at the smallest shells by viscosity
$\nu=10^{-7}$. With these parameters, our simulations correspond to
a Reynolds number $Re\simeq 10^{8}$.
For each realization, Lagrangian structure functions 
are computed from the Lagrangian velocity (\ref{eq:6}) up to the
large-scale time $\tau_{0}$. Average is then taken over $10^{5}$
independent realizations.

\begin{figure}[htb]
\epsfxsize=8.5cm
\centerline{\epsfbox{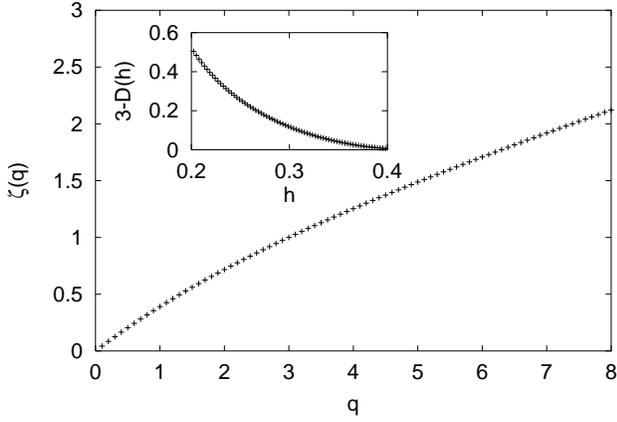}}
\narrowtext
\caption{Shell model Eulerian structure function scaling exponents
$\zeta(q)$ determined from over the statistics of $10^5$
independent configurations. In the inset we plot the 
codimension $3-D(h)$ as determined from numerically solving
the Legendre transform.}
\label{fig1}
\end{figure}

In Fig.~\ref{fig1} we plot the set of numerically determined 
Eulerian structure function scaling exponent $\zeta(q)$ together with
the fractal dimension $D(h)$ reconstructed by means of the Legendre
transform. We observe strong intermittency in velocity statistics
with scaling exponent which deviates from Kolmogorov prediction.
The scaling exponents are not universal with respect to the 
particular shell model. Model (\ref{eq:5}) gives a set of exponents
which are a little more intermittent than, but not far from, the
experimentally observed exponents \cite{Frisch95}:
$\zeta(2)\simeq 0.72$, $\zeta(4)\simeq 1.25$, $\zeta(6)\simeq 1.71$.
We thus expect that the values of $\xi(p)$ obtained from (\ref{eq:4})
using the $D(h)$ of Fig.~\ref{fig1} will be directly comparable with real
experimental data.

\begin{figure}[htb]
\epsfxsize=8.5cm
\centerline{\epsfbox{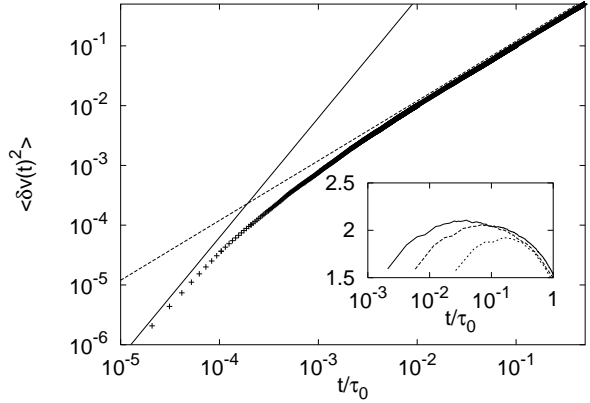}}
\narrowtext
\caption{Second-order Lagrangian structure function
$\langle \delta v(t) ^2 \rangle$ as a function of time
delay $t$ for the simulation at $Re=10^{8}$.
Continuous line is the ballistic behavior
$t^2$ at short time. Dashed line represents the
linear growth (\ref{eq:1}).
Inset: $\langle \delta v(t) ^2 \rangle$
compensated with the dimensional prediction $\varepsilon t$
at $Re=10^{8}$ (continuous line), $Re=2\times 10^{6}$ (dashed line)
and $Re=10^{5}$ (dotted line).}
\label{fig2}
\vspace{2cm}
\end{figure}

Figure \ref{fig2} shows the second-order Lagrangian structure 
function (\ref{eq:1}) as a function of time. The linear behavior
is evident (see the inset) even if a long crossover from 
the ballistic scaling at short time 
$\langle \delta v(t) ^2 \rangle \sim t^2$ is present. 
In spite of the very high Reynolds numbers achievable in the shell model,
the extension of the temporal scaling (\ref{eq:2}) is still moderate.
For a comparison with the available
experimental data, in the inset we also plot the result obtained
from two simulations at lower resolution, with $Re \simeq 2 \times 10^{6}$
and $Re \simeq 10^{5}$. In the latter case almost no scaling 
range is observable. 
Despite these limitations, we will see that high $Re$ simulations
allow the determination of the Lagrangian scaling exponents with
good accuracy.

The long crossover in Fig.~\ref{fig2} can be understood in terms
of intermediate dissipative range as a consequence of the fluctuating
dissipative scale \cite{Frisch95,FV91}. The smallest time at which
one can expect scaling (\ref{eq:1}) is the Kolmogorov time
$\tau_{\eta} \sim \tau_{0} Re^{-(1-h)/(1+h)}$ which fluctuates with
$h$. A demonstration of the effects induced by intermittency 
is given by considering a non-intermittent Gaussian model. 

\begin{figure}[htb]
\epsfxsize=8.5cm
\centerline{\epsfbox{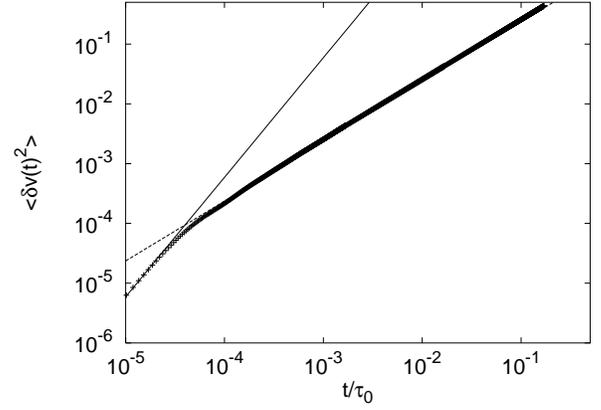}}
\narrowtext
\caption{Second-order Lagrangian structure function
$\langle \delta v(t) ^2 \rangle$ as a function of time
delay $t$ for the equilibrium Gaussian model.
Continuous line is the ballistic behavior
$t^2$ at short time. Dashed line represents the
linear growth (\ref{eq:1}).}
\label{fig3}
\end{figure}

Setting $f_{n}=\nu=0$, (\ref{eq:5}) becomes a conservative system
with two conserved quantitities which depends on the value of the
$\delta$ \cite{BJPV98}. In statistical stationary condition,
the model shows equipartition of the conserved quantities among the
shell, in agreement with statistical mechanics prediction
\cite{ABCFPV94}.
For $\delta=1+2^{-2/3}$ the equipartition state leads at small scales
to Kolmogorov scaling $\langle |u_{n}|^2 \rangle \sim \ell_{n}^{2/3}$
with Gaussian statistics. In Fig.~\ref{fig3} we plot the second-order
Lagrangian structure function (\ref{eq:1}) for the Gaussian model.
Both the ballistic and the diffusive scaling is clearly observable
and the crossover is strongly reduced with respect to Fig.~\ref{fig2}.

\begin{figure}[htb]
\epsfxsize=8.5cm
\centerline{\epsfbox{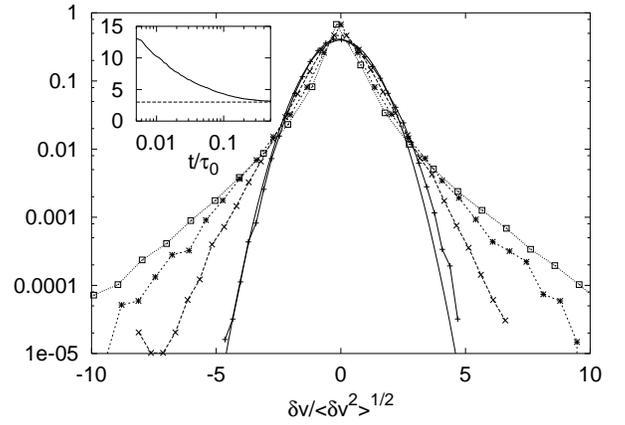}}
\narrowtext
\caption{Probability density functions of velocity differences 
$\delta v(t)$ normalized with the variance at time lags
$t/\tau_0=0.002 (\Box), 0.01 (*), 0.06 (\times), 0.35 (+)$.
Continuous line represents a Gaussian. Inset: flatness
$F=\langle \delta v(t)^4 \rangle / \langle \delta v(t)^2 \rangle^2$
as function of time and Gaussian value $F=3$ (dashed line).}
\label{fig4}
\end{figure}

In Fig.~\ref{fig4} we plot the probability density functions
of $\delta v(t)$ computed at different $t$ in the linear
scaling range of Fig.~\ref{fig2} rescaled with their variances.
The form of the pdf varies continuously from almost Gaussian
at large time ($t \sim \tau_0$) to the development of 
stretched exponential tails at short times, similar to 
what observed in laboratory experiments \cite{MMMP01}.
Flatness $F$ grows from Gaussian value $F=3$ up to 
$F \simeq 20$ at smallest times.
This is an indication of Lagrangian intermittency,
in the sense that the Lagrangian
statistics cannot be described in term of a single scaling
exponent.

\begin{figure}[htb]
\epsfxsize=8.5cm
\centerline{\epsfbox{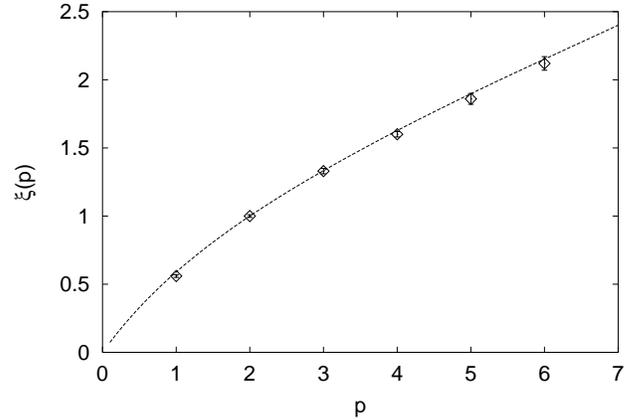}}
\narrowtext
\caption{Lagrangian structure function scaling exponents
$\xi(p)$ numerically determined by a best fit on (\ref{eq:2}).
The line represents the multifractal prediction (\ref{eq:4})
with $D(h)$ obtained from Fig.~\ref{fig1}.}
\label{fig5}
\end{figure}

In Fig.~\ref{fig5} we plot the set of Lagrangian scaling exponents
$\xi(p)$ obtained from a direct fit of temporal structure functions.
The nonlinear behavior in $p$ confirms the presence of Lagrangian
intermittency already observed from the pdf.
We present the result for moments up to $p=6$ which approximatively
corresponds, from (\ref{eq:4}), 
to Eulerian structure function of order $q=8$. In this sense
temporal structure functions are more intermittent.
Figure~\ref{fig5} shows that the agreement with the multifractal
prediction (\ref{eq:4}) is very good up to the moment achievable
with our statistics.
What is even more remarkable is that our prediction is very close
to experimentally determined exponents. For example
we find $\xi(3)\simeq 1.31$, $\xi(4)\simeq 1.58$, $\xi(5)\simeq 1.85$,
while the experimental data give \cite{MMMP01} $\xi_{exp}(3)=1.34 \pm 0.02$,
$\xi_{exp}(4)=1.56 \pm 0.06$ and $\xi_{exp}(5)=1.8 \pm 0.2$.

In this work we have investigated the statistical properties of
single particle Lagrangian velocity in fully developed turbulence.
A prediction for intermittent scaling exponents of Lagrangian structure
function is given within the multifractal framework.
Very high Reynolds number simulations in shell model confirm
the multifractal prediction, even if rather small scaling ranges
are observed. In particular, our simulations show that at 
experimental Reynolds number presently available almost no
scaling is observable. The reduction of the scaling range in
Lagrangian statistics is interpreted as an effect of the intermediate
dissipative range.  A Gaussian, non-intermittent version of the shell
model confirms this interpretation.
An important consequence of our findings is that usual models
of particle dispersion, based on stochastic model \cite{Thomson87},
are incorrect in a fundamental sense, and one should take into
account the modifications due to intermittency.

This work was supported by MIUR-Cofin2001 contract 2001023848.
We acknowledge the allocation of computer resources from INFM
``Progetto Calcolo Parallelo''.

\newpage

\end{multicols} 
\end{document}